\documentclass{jltp}
\usepackage{graphicx,amssymb}

\title{Thermally Induced Losses in Ultra-Cold Atoms Magnetically Trapped Near Room-Temperature Surfaces}

\author{D.~M. Harber, J.~M. McGuirk$^*$, J.~M. Obrecht, and E.~A. Cornell$^*$}

\address{JILA, National Institute of Standards and Technology \\
and University of Colorado, and Department of Physics, \\ 
University of Colorado, Boulder, Colorado 80309-0440, USA\\
$^*$Quantum Physics Division, National Institute of Standards and Technology.}

\runninghead{D.~M. Harber, J.~M. McGuirk, J.~M. Obrecht, and E.~A. Cornell}{Thermally Induced Losses in Ultra-Cold Atoms Near Surfaces}

\begin{document}

\maketitle

\begin{abstract}
We have measured magnetic trap lifetimes of ultra-cold $^{87}$Rb atoms at distances of 5-1000 $\mu$m from surfaces of conducting metals with varying resistivity.  Good agreement is found with a theoretical model for losses arising from near-field magnetic thermal noise, confirming the complications associated with holding trapped atoms close to conducting surfaces.  A dielectric surface (silicon) was found in contrast to be so benign that we are able to evaporatively cool atoms to a Bose-Einstein condensate by using the surface to selectively adsorb higher energy atoms.


\end{abstract}


\section{INTRODUCTION}

Surface microtraps have recently seen increasing popularity in the manipulation of neutral ultracold atoms.\cite{reichel2001,prentiss1998,Zimmermann2001,pritchard2002,schmied2003,hinds2003,prentiss2002,prentiss1999}  Fabrication techniques developed for microelectronics may be used to create intricate layouts of current-carrying wires on surface substrates.\cite{prentiss2002}  Large magnetic field gradients, and thus trapping forces, can be generated in close proximity to the current-carrying wires ($\sim$10-100 $\mu$m) with relatively modest currents ($\sim$1~Amp).  Additionally, the flexibility of the fabrication techniques allows the experimentalist a great deal of control over the trapping potentials; atom guiding and interferometry,\cite{pritchard2002,schmied2003,cornell2001,schmied2002} double well potentials,\cite{schmied2003,hansch2001} and even atom-cavity coupling schemes\cite{hinds2003b} have been proposed or are currently being attempted with surface microtraps.

Unfortunately, there exist several roadblocks and detours on this promising avenue of research.  Atom cloud fragmentation has been observed when the trapped atoms are brought in close proximity to the current-carrying wires.\cite{zimm2002,ketterle2003}  This effect is believed to be caused by current path deviations within the wires that produce magnetic field deviations parallel to the wires, and thus corrugate the potential.\cite{zimm2002b}  Additionally, increased atom loss and heating have been observed when trapped atoms approach the guiding wire.\cite{zimm2002,ketterle2003}  The atom loss has been primarily attributed to radio-frequency (rf) noise in the wires, caused by either the current supply or antenna pickup effects.  The rf noise is radiated from the wires, and, if at the Larmor frequency $\nu_L$, can induce the atoms to make spin flip transitions to untrapped Zeeman sublevels.  Likewise, current noise at harmonics of the trap frequencies can induce heating.

The effects mentioned above are essentially technical problems related to the current-carrying wires on the surface; however in a recent publication by Jones et al.\cite{hinds2003} a more fundamental loss mechanism was documented.  As the atoms were moved near the current carrying wires, increased atom loss was seen.  It was shown that the distance scaling behavior did not match that expected from a radiating wire, and instead appeared consistent with fields emanating from the solid.  This effect, predicted by Henkel et al.,\cite{wilkens1999} is due to thermal current fluctuations present in conducting solids.  These current fluctuations, which increase with decreasing resistivity of the solid, induce magnetic field fluctuations that increase in strength close to the surface of the solid.  If an atom is sufficiently close to a conducting surface, magnetic field fluctuations at the Larmor frequency can drive Zeeman spin flip transitions to untrapped states, similar to the technical loss mechanism described above.

In this report, we present measurements of surface loss in which a conventional magnetic `macro-trap' is used for confinement, and supplementary magnetic fields are used to move the atoms close to the surfaces.  This configuration has allowed us to measure loss rates over a number of different surfaces with varying resistivity.  Additionally, loss measurements were made in traps with different bias fields, and thus different Larmor frequencies.  We compare our results to theory, and find good quantitative agreement.  Finally, we demonstrate evaporative cooling using a surface (instead of a rf ``knife") to perform position-selective atom removal.  We are able to traverse our entire evaporation trajectory using this evaporation technique, and Bose-condensates containing up to $5\times10^5$ atoms can be formed. 

Use of a trapping potential independent of the surface provides several advantages.  First, the trapping potential remains constant as the atoms approach the surface, which is considerably more difficult to achieve in a microtrap, allowing loss measurements at a fixed Larmor frequency to be made.  Second, since we do not rely on wires on our surface, we may work over surfaces with a variety of resistivities, and we can prepare those surfaces with atom-surface measurements in mind. 

\section{EXPERIMENT}

\begin{figure}
\centerline{\includegraphics[width=4in]{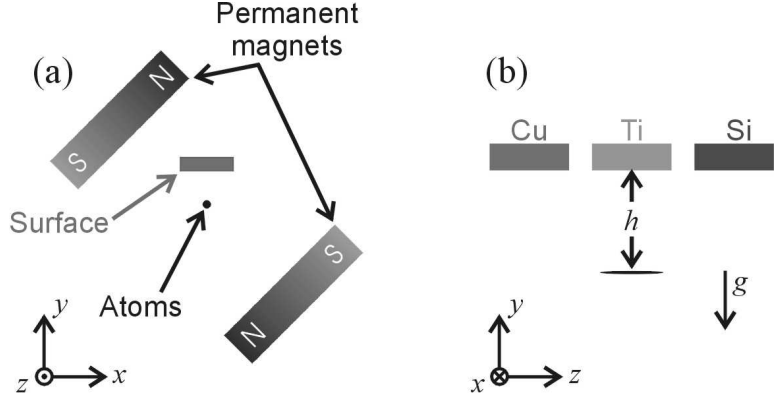}}
\caption{(a) Axial view of trap and surface geometry. (b) Side view of atoms and surface geometry.  Adding a supplementary magnetic field moves atoms perpendicular to surface ($\pm y$ direction); horizontal translation of I-P trap moves atoms from surface to surface ($\pm z$ direction).}  
\label{fig:diagram}
\end{figure}

The experimental apparatus, described in detail in Ref. \onlinecite{lewando2003}, begins with a vapor cell magneto-optical trap that collects $\sim$$10^{10}$ $^{87}$Rb atoms.  The atoms are then optically pumped into the $|F=1\rangle$ state, and a quadrupole trap is turned on, trapping atoms in the $|F=1,m_f=-1\rangle$ state.  The quadrupole coils, which are mounted on a linear servo-motor controlled track, are then moved 40 cm to transport the atoms to the Ioffe-Pritchard (I-P) trap.  The I-P trap is a hybrid design that uses a pair of permanent magnets to provide a radial gradient of 450 gauss/cm.  Axial confinement and control of the bias field are provided by two pairs of electromagnetic coils.  The I-P trap is $\sim$6 cm long in the axial direction, and has a inner diameter of $\sim$2 cm; this is significantly larger than both atom cloud sizes and atom-surface separations.  At a typical bias field of 3.2 gauss the $|F=1,m_f=-1\rangle$ atoms experience trap frequencies of 7 Hz and 230 Hz in the axial and radial directions, respectively.  Once the atoms have been transferred to the I-P trap they have a temperature of $\sim$500 $\mu$K; further cooling is accomplished by rf-evaporation, and condensates with up to $5\times10^5$ atoms can be created.  

Once a normal cloud or Bose condensate has been prepared, the atoms are moved towards the surface for a measurement.  This is accomplished by the application of a vertical magnetic field, or push field, that shifts the trap center towards the surface ($y$ direction, Fig. 1).  The unperturbed trap center is $\sim$1 mm from the surface, so a vertical magnetic field of $\sim$45 gauss must be applied to move the atoms to the surface.  Our apparatus also allows multiple surface samples to reside in the chamber (Fig. 1b); to move from surface to surface the magnetic trap is translated $\pm4.5$ mm in the $z$ direction.  For this work we used $\sim$1 mm thick polished surfaces of silicon, copper, and titanium.

Imaging of the atoms following a measurement was accomplished in one of several ways.  First, the atoms can simply be moved back to the unperturbed trap center after a measurement, and imaged through absorption.\cite{imaging}  Second, the push field can remain on during imaging; thus the atoms appear in the image vertically displaced from the initial trap center.  To image the atoms near the surface we use a technique performed by S. Schneider et al.\cite{schmied2003}  Rather than align the beam parallel to the surface, the beam is aligned with a slight grazing incidence, $1.7^\circ$ in our case.  If atoms are then imaged sufficiently near the surface ($<100$ $\mu$m) a second image reflected by the surface will appear.  Measurement of the vertical position of the two images allows a determination of both the atom-surface distance and surface position to be made.  Using this technique we are able to measure the atom-surface separation to within an uncertainty of 1 $\mu$m.  Additionally, this technique can be used to measure the angle of the cloud with respect to the surface; we find that the axis of the cloud is parallel to within $0.3^\circ$ of the surface, which corresponds to $\sim$1 $\mu$m additional atom-surface uncertainty in a typical normal cloud and $\sim$0.3 $\mu$m in a condensate.

\begin{figure}
\centerline{\includegraphics[width=2.5in]{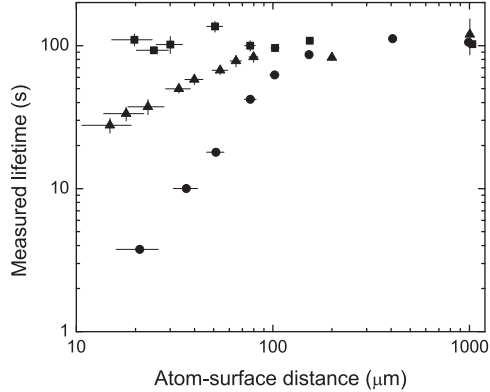}}
\caption{Lifetimes of non-condensed (normal) atoms near surfaces of copper({\large$\bullet$}), titanium ($\blacktriangle$) and silicon ({\tiny$\blacksquare$}); $\nu_L = 1.80$ MHz for these measurements.  The finite width of the atom clouds becomes significant at close distances to the surface, so the radial width of the atom clouds is reflected by the $x$-error bars.}
\label{fig:fig2}
\end{figure}

Trap lifetime measurements were made with both normal clouds and condensates.  To make a trap lifetime measurement either a normal cloud at 300-500 nK containing $\sim$$5\times10^5$ atoms or a condensate containing $\sim$$4\times10^4$ atoms with a negligible normal component is prepared in a trap with an axial bias field $B_z$ of 3.2 gauss.  Then, over 200-300 ms, the trap center is moved to the desired distance from the surface, and simultaneously the trap bias field is ramped to the desired value, typically 2.57 gauss, corresponding to a Larmor frequency of 1.80(1) MHz.  A low power rf-shield is then turned on $\sim$400 kHz above the trap bottom for a normal cloud, or $\sim$80 kHz for a condensate, for the duration of the hold time (between 1 ms and 300 s).  This is done to minimize the heating effects of the Oort cloud.\cite{wieman1999}  Finally, the trap center is ramped back to the initial value, and the imaging procedure is performed.

\section{RESULTS}

These measurements were performed over three different surface materials: copper, titanium, and silicon, with corresponding resistivity of $1.67(8)\times10^{-8}$ $\Omega$-m, $4.88(24)\times10^{-7}$ $\Omega$-m, and $>$$1$ $\Omega$-m\cite{resistivity} and surface dimensions of $\sim$8$\times$1$\times$3 mm ($x$$\times$$y$$\times$$z$, see Fig. 1).  The measured normal cloud lifetimes as a function of distance $h$ to the surface are shown in Fig. 2.  A number of trends are immediately evident.  First, the lifetime over the copper surface drops from the background gas limited rate of $\sim$120 s to $\sim$4 s in only 200 $\mu$m; this is clearly not a subtle effect.  Second, the lifetimes are significantly longer over titanium, which has a resistivity 28 times larger than copper, and over silicon no statistically significant lifetime reduction was observed for atom-surface separations $>$10 $\mu$m.

For atom-surface separations in the 100 $\mu$m range, trap lifetimes over copper and titanium become limited to $\sim$120 s due to collisions with the background gas.  In order to examine the trap loss due only to surface effects, the background gas-limited lifetime $\tau_{BG}$ can be subtracted from the measured lifetime $\tau_0$ as

\begin{equation}
\tau = (1/\tau_0 - 1/\tau_{BG})^{-1},
\end{equation} 

\noindent to yield the surface-limited lifetime $\tau$.  The surface-limited lifetimes for copper, titanium and silicon are shown in Fig. 3(a).  Additionally, lifetime measurements were made over copper in a trap with a Larmor frequency of 6.24(1) MHz to study Larmor frequency dependence, and the surface-limited lifetimes for both 1.80 and 6.24 MHz Larmor frequencies are shown in Fig. 3(b).

Theoretical predictions for loss above the surfaces were made following theory derived by Henkel et al.\cite{wilkens1999}  Assuming atom loss occurs via the $|F=1,m_f=-1\rangle \rightarrow |F=1,m_f=0\rangle$ transition,\cite{hyperfineloss} we numerically integrate the integrals in Eq. (22) of Ref.~\onlinecite{wilkens1999} to get the predicted value of the surface-limited $\tau$.  The suggested closed-form asymptotic interpolation, Eq. (23) of Ref.~\onlinecite{wilkens1999}, is plotted as a dotted line in Fig 3(a).  It provides the correct trends, but is not very accurate in the region $h\sim\delta$ where $\delta$, the skin depth, is expressed as

\begin{equation}
\delta = \sqrt{\frac{2\varepsilon_0c^2\rho}{2\pi\nu_L}},
\end{equation}

\noindent where $\varepsilon_0$ is the permittivity of free space and $\rho$ is the resistivity of the surface.

\begin{figure}
\centerline{\includegraphics[width=4.9in]{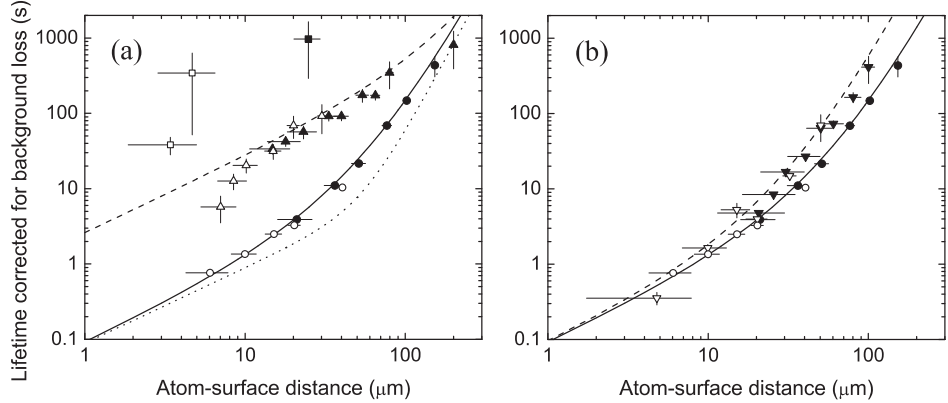}}
\caption{Inferred lifetimes near surfaces after subtraction of background loss.  (a) Resistivity dependence: copper ({\large$\bullet$}), titanium ($\blacktriangle$) and silicon ({\tiny$\blacksquare$}) at $\nu_L=1.80$ MHz.  Open $\vartriangle$ and filled $\blacktriangle$ symbols represent measurements made with condensates and normal clouds respectively.  Solid (Dashed) line indicates lifetimes predicted for copper (titanium).  The solid and dashed theory lines are based on a numerical integration of Eq. (22) of Ref.~\onlinecite{wilkens1999}.  The dotted line, plotted for copper, shows the closed-form interpolation suggested in Ref.~\onlinecite{wilkens1999}. (b) Larmor frequency dependence near copper: $\nu_L=1.80$ MHz ({\large$\bullet$}) and $\nu_L=6.24$ MHz ($\blacktriangledown$).  Again, open and filled symbols represent condensate and normal clouds respectively. Solid (Dashed) line indicates lifetimes predicted for $\nu_L$ = 1.80 MHz (6.24 MHz).}
\label{fig:fig3}
\end{figure}

We see good agreement between the measured and predicted lifetimes for both the copper at 1.80 and 6.24 MHz Larmor frequencies ($\delta = 49$ $\mu$m and $26$ $\mu$m respectively), as well as the titanium at 1.80 MHz Larmor frequency ($\delta = 262$ $\mu$m).  At the shortest atom-surface distances, the measured lifetimes appear to drop anomalously rapidly.  This effect was also seen over silicon, and we suspect that near this length scale the Casimir-Polder force\cite{CasimirPolder} begins to become significant, and can possibly be invoked to explain increased loss.\cite{vuletic}

The leading systematic effects in this measurement are three-body loss, loss of atoms that tunnel into the surface, and the finite sizes of the atom clouds.  Three-body loss, independent of distance to the surface, occurs in normal clouds and condensates, but is much larger in condensates (three-body loss limits condensate lifetimes to $\sim$40 s).  Due to its lack of dependence of distance to the surface, three-body loss can be removed with $\tau_{BG}$.  Loss of atoms that strike the surface, or mechanical shaving, is dependent on distance to the surface and the width of the atom cloud.  Working with normal clouds and condensates, with typical $1/e$ and Thomas-Fermi radii of 4.2 $\mu$m and 1.5 $\mu$m, respectively, tests for this effect.  Good agreement is seen between surface-limited lifetimes measured with condensates and normal clouds, so we feel neither of these systematics is problematic.

\subsection{Surface Evaporation}

Alternatively, the loss of atoms that strike the surface, or mechanical shaving, can be used constructively.  By using a surface for position selective removal of atoms, and thus energy selective, one effectively obtains an evaporation knife.\cite{reichel1999}  Further, magnetic trap lifetimes over high resistivity materials, such as silicon, are essentially only limited by the background gas collision rate.  One could imagine using a high resistivity material as an evaporation knife in the place of rf.  Much like rf evaporation, atoms with trajectories that bring them furthest from the trap center, i.e. the most energetic atoms, are removed.  Unlike rf evaporation, this evaporation occurs only along one dimension, possibly limiting its efficiency.  On the other hand, along this dimension the surface acts as a nearly ideal evaporation surface.  As an evaporative knife, the dielectric surface may in effect be ``sharper," removing the high-energy atoms with more certainty, while leaving the low-energy atoms less perturbed than an rf knife.

\begin{figure}
\centerline{\includegraphics[width=3.5in]{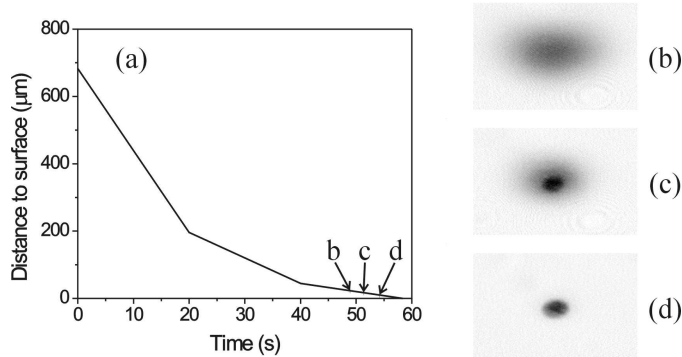}}
\caption{Surface evaporation to Bose-condensation: (a) Trap center-surface distance trajectory for silicon surface evaporation.  Images showing progressively closer evaporation endpoints and the appearance of the Bose-condensate: final atom-surface separations correspond to (b) 23 $\mu$m (normal cloud), (c) 17 $\mu$m (partially condensed), and (d) 10 $\mu$m (nearly pure condensate).}
\label{fig:fig4}
\end{figure}

To demonstrate the effectiveness of this technique we load the I-P trap as usual; however we do not apply any rf for evaporation.  Instead we apply a series of linear ramps of the vertical magnetic field, which moves the trap center towards the silicon surface (see Fig. 4).  The evaporation
trajectory is begun with the trap center 700 $\mu$m from the surface (our initial temperature is $\sim$500 $\mu$K).  Over 50 seconds the trap center is ramped gradually towards the surface, and at a final separation from the trap center to the surface of 10 $\mu$m we form a Bose-condensate of $\sim$5$ \times 10^5$ atoms.  This is comparable to the size of condensates that can be created through rf evaporation, leading one to conclude that the efficiency of the mechanical evaporation is similar to that of rf evaporation.\cite{anderson}

This technique of evaporation may be particularly well suited for continuous condensation schemes.\cite{odelin2002}  In this case it may be quite difficult to apply spatially varying rf.  Instead, by sending the cloud close to a surface, or more specifically having a spatially variable atom-surface separation in an atom guide, evaporation to condensation may be performed ``on the fly."

\section{CONCLUSION}

In conclusion, we have performed a series of magnetic trap lifetimes measurements over 5-1000 $\mu$m distance scales exploring the loss dependence on surface resistivity and atomic Larmor frequencies.  We find good agreement with theoretical predictions.  This confirms the problems associated with working at short atom-surface separations over low resistivity metals.  

There may exist several possible solutions to this problem.  The simplest but possibly the most undesirable is simply to restrict atom-surface separations to separations larger than the skin depth $\delta$, which negates some of the advantages that microtraps provide.  Second, thin wires, with thickness $\ll\delta$, can be used.  This significantly can reduce the fluctuating fields generated\cite{hinds2003}; however power dissipation becomes increasingly problematic as wires become small.  Similarly, wires made from higher resistivity materials can be used, but again power dissipation is a problem.  Cooling of the microtrap surface and wires has also been proposed\cite{schmied2003b}; however for many metals the resistivity scales as the temperature, and thus for regions where $h\ll\delta$ the lifetime will not be affected.  Cooling of the surface, however, will decrease the skin depth, and consequently increase lifetimes outside of that skin depth.  The use of superconducting wires may skirt these problems, but a detailed discussion is outside the scope of this publication.  Alternatively, potentials may be generated without the use of current-carrying wires; microtraps using permanent magnet structures,\cite{hinds2003,prentiss2002} electrostatic potentials,\cite{schmied2003c} and microoptics\cite{ertmer2002} have been demonstrated.

Finally, we have demonstrated an evaporation technique where a silicon surface is used in the place of rf.  This technique may be particulary useful in situations where the application of rf may be problematic, such as continuous condensate systems.

\section*{ACKNOWLEDGMENTS}

We acknowledge useful conversations with the other members of the JILA BEC collaboration.  This work was supported by grants from the NSF and NIST.  This material is based upon work supported under a National Science Foundation Graduate Research Fellowship.


\begin{thebibliography}{9}
\bibitem{imaging}
Imaging is performed by transferring the atoms from $|F=1,m_f=-1\rangle \rightarrow |F=2,m_f=-2\rangle$ through adiabatic rapid passage.  The antitrapped atoms rapidly expand, and are imaged through absorption with a 40 $\mu\mbox{s}$ pulse of $5\mbox{S}_{1/2}$ $|2,-2\rangle \rightarrow 5\mbox{P}_{3/2}$ $|3,-3\rangle$ light.
\bibitem{resistivity}
Four-wire resistivity measurements were made using similar samples of copper ($99.9999\%$ purity), and titanium ($99.7\%$ purity) to what were used for the lifetime measurements; the purity of the silicon sample was $99.999\%$.
\bibitem{hyperfineloss}
At distances within 10 $\mu$m of titanium, microwave transitions to the $|F=2\rangle$ levels may begin to contribute significantly to the loss as well; this is not accounted for in the plotted theory curves.
\end{thebibliography}
\end{document}